# Edge and corner states in 2D non-Abelian topological insulators from an eigenvector frame rotation perspective

Tianshu Jiang, Ruo-Yang Zhang, Qinghua Guo, Biao Yang and C. T. Chan*

*Department of Physics and Center for Metamaterials Research, Hong Kong University of Science and Technology, Clear Water Bay, Hong Kong, China*

* Correspondence address: phchan@ust.hk

**Abstract**

We propose the concept of 2D non-Abelian topological insulator which can explain the energy distributions of the edge states and corner states in systems with parity-time symmetry. From the viewpoint of non-Abelian band topology, we establish the constraints on the 2D Zak phase and polarization. We demonstrate that the corner states in some 2D systems can be explained as the boundary mode of the 1D edge states arising from the multi-band non-Abelian topology of the system. We also propose the use of off-diagonal Berry phase as complementary information to assist the prediction of edge states in non-Abelian topological insulators. Our work provides an alternative approach to study edge and corner modes and this idea can be extended to 3D systems.

## I.   INTRODUCTION

Higher-order topological insulators (HOTIs) have attracted a lot of interest in recent years. The most notable feature of HOTI is the *d-m* dimension edge states in the *d* dimension topological insulator with codimension $m > 1$. Typical examples are the 0D corner states in the 2D insulators [1-19]. The HOTIs can have different classifications and can be realized in different forms, such as multipole insulators [2-4], breathing Kagome lattices [5] and generalized 2D/3D Su-Schrieffer-Heeger (SSH) lattices [6-10]. The widely used theories of the bulk-boundary correspondence include the multipole moments [2,3], dipole polarization [6], filling anomaly [7,11] and real Chern number [12]. However, these theories are based on the topological theory that classify each band or each band gap individually, but the HOTIs are usually multi-band systems. In some experimental results and numerical simulations, the energy locations of the first-order edge states and second-order corner states can appear either in the band gaps or in the bulk bands [13-19] and there are cases where the location of the corner modes cannot be predicted by previous theories such as the 2D bulk polarization [6]. In this work, we will see that some of these phenomena can be explained using a multiband approach.

While conventional topological band theories focus on the topological character of an individual band or band gap, there are alternative approaches that consider multiple bands simultaneously. A recently proposed approach considers the frame rotation of the eigenvector frame of several states and the topological invariants are elements of the quaternion group which are matrix-like quantities (not numbers) that describe the rotation, and this approach is called non-Abelian band theory because rotations are typically non-commutative [20-27]. In this approach, the nodal line configurations in momentum space are characterized by non-Abelian groups as the fundamental homotopy groups of the Hamiltonian space in multi-band parity-time reversal (PT) symmetrical systems. Based on this theory, a quasi-1D non-Abelian topological insulator with multiple states per unit cell has been experimentally realized and the bulk-edge correspondence has been studied [22,24]. We can always extend the 1D periodic model to higher dimensions so that the 1D Brillouin zone (BZ) corresponds to a closed loop in a higher-dimension momentum space. It can be shown that if the eigenstate frames on 1D loop have a non-trivial quantized rotation, the loop must enclose nodal lines (degeneracies) and a non-Abelian charge (quaternion or generalized quaternion) can be assigned to topologically characterize the encircled nodal lines. The bulk-edge correspondence between the bulk non-Abelian topological charges and the edge states are more subtle than Abelian quantities such as Chern numbers. For example, in 1D non-Abelian insulators, there are configurations in which the edge states can be indeterminate even if their quaternion charges are specified [20,22,24]. In particular, in a 3-band system, the exact location of edge states of the topological quaternion charge $-1$ and the number of edge states cannot be determined a priori. Similar to the charge $-1$ in a 3-band system, the generalized quaternion charge charges $q_{1234}$ and $-q_{1234}$ in a 4-band system can also have many possibilities in the edge state distributions in the energy spectra. These topologically allowed configurations can be continuously transformed to each other without bandgap closure for a given non-Abelian charge. The bulk edge correspondence is more direct in traditional topological band theories in which the topological invariants are integers and as such, they can be mapped directly to the number of edge state modes. In non-Abelian theories, complexity arises because non-Abelian topological charges are not integers and so the correspondence with the number of edge modes is indirect. Such a complexity explains the fairly complicated energy distributions of edge and corner states in some HOTIs, where a multiband description is needed. However, other topological indicators such as the off-diagonal Berry phase (to be defined below) can provide additional information to assist the establishment of the bulk-edge correspondence.

In this study, we propose 2D topological insulators from a multiband perspective. We will consider 3-band and 4-band models as prototypical examples. We find that the generalized 2D SSH lattices in the HOTI family can be classified as a 2D non-Abelian topological insulator and the corresponding edge state distributions can be explained by the non-Abelian band topology. In addition to the non-Abelian quaternion charge, we show that the off-diagonal Berry phase (OBP) can provide complementary information and help us to predict the edge state distributions. Furthermore, we show that the higher-order corner states in these systems can be viewed as the 0D edge state of the 1D edge state. We note that some recent papers [13,14] also argued that the corner states in some systems arise from edge-localized bands. According to this viewpoint, the energy locations of corner states can be explained and predicted by calculating the Zak phase winding of the edge states. We will also discuss more generalized 2D lattices that do not meet the definition of 2D non-Abelian topological insulators.

## II. 2D NON-ABELIAN TOPOLOGICAL INSULATOR

In this section, we discuss using the notion of eigenvector frame rotation to define a 2D non-Abelian topological insulator and discuss its properties.

### A. From 1D to 2D non-Abelian topological insulator

The concept of non-Abelian band topology has proved useful in describing the admissible nodal line configurations in the 3D Brillouin zone (BZ) [20,25-27] and Dirac nodal point braiding in the 2D BZ [21,23] in PT symmetrical multi-band systems. The corresponding fundamental homotopy groups of the order-parameter Hamiltonian spaces are non-Abelian groups. The non-Abelian charges are defined by the eigenvector frame rotation along a closed loop in the BZ.

We will first review the 1D non-Abelian topological insulator before we move on to 2D. Let us consider a 1D periodic system with 3 atoms per unit cell. The system has PT symmetry, and as such, the Hamiltonian matrix can be gauged to be real in k-space with a proper choice of basis and all eigenvectors are real. All 3 bands are required to be gapped, so that they can be labeled sequentially as 1,2,3. For simplicity, we can examine a flat band model with a Hamiltonian as $H(k_x) = R(k_x) diag(-1,0,1) R(k_x)^T$ with $R(k_x) = \exp[\vec{\beta}(k_x) \cdot \vec{L}]$, where $(L_i)_{jk} = -\epsilon_{ijk}$, $\epsilon_{ijk}$ is the fully antisymmetric tensor, and $\vec{\beta}(k) = \phi \hat{n}$ is the product of rotation axis $\hat{n}(k)$ and rotation angle $\phi(k)$. The eigenstate frame rotation in the 1D three-band system can be illustrated as shown in Figs. 1(a)-1(c). The band dispersion (flat bands) are shown in Fig. 1(c). In Fig. 1(a), we plot the frame of eigenstates along the 1D first BZ. The director of each eigenvector is indicated by a small bar, with color corresponding to the bands labelled in Fig. 1(c). We note that the eigenvector frame is rotating, as $k_x$ moves across the 1D BZ. Frame rotations in 3D can be described using the quaternion group. For this flat band Hamiltonian, the elements of the quaternion group $+i$, $+j$, $+k$ correspond to $R(k_x) = \exp(\frac{k_x+\pi}{2} L_{x,y,z})$ respectively. Here, we set $R(k_x) = \exp(\frac{k_x+\pi}{2} L_z)$. The eigenstate can be expressed by a vector with 3 real number components under a suitable basis. The red, green and blue sticks correspond to the eigenstates of the 1st, 2nd and 3rd bands with blue sticks being perpendicular to the page (see the band structure in Fig. 1(c)), respectively. They compose a frame because the eigenstates of the three bands are orthogonal to each other. The frame rotates an $\pi$ angle around the blue axis (3rd band) with the Bloch $k_x$ running from $-\pi$ to $\pi$. According to Ref. [[20]], the quaternion charge classes $\{\pm i\}$, $\{\pm j\}$, $\{\pm k\}$ correspond respectively to the $\pm\pi$ eigenstate rotations around the axis of the 1st, 2nd, 3rd bands, and the class $\{-1\}$ corresponds to a $2\pi$ rotation around any axis, and $\{+1\}$ is the topologically trivial configuration without frame rotations. Using such a convention, Fig. 1a indicates that the model 1D system can be labelled with quaternion charge $+k$. Generally, we can numerically calculate the non-Abelian charge which is introduced in the section 1 of Ref. [28].

We can intuitively understand the relationship between the nodal lines and the 1D non-Abelian topological insulator from a higher dimensional picture. We consider a 3D system with nodal line (degeneracy) configurations as shown in Fig. 1(b). The blue nodal line denotes a nodal line (degeneracy points between the 1st and 2nd bands) that can be characterized by the quaternion "$k$" in parameter space and the orientation (the arrow) represents its sign. Now we choose a closed loop (black loop) in this 3D parameter space that encircles the blue line. The Hamiltonians on this loop define a 1D subsystem if we view the momentum along the loop as the Bloch $k_x$. We label the nodal line as a "$+k$" nodal line as the rotation of the eigen-frame on the loop encircling the line is characterized by the non-Abelian charge $+k$ [22]. Therefore, we can view the 1D model in Fig. 1(a) as a 1D subsystem in the 3D space which constitutes a loop enclosing a nodal line (Fig. 1(b)).

Now we generalize the concept of 1D non-Abelian topological insulator to the 2D system. The 2D system is required to be PT-symmetric with 3 or more gapped bands. Here we consider a 2D model Hamiltonian expressed as $H(k_x, k_y) = R_2(k_y)R_1(k_x)diag(-1,0,1)R_1(k_x)^T R_2(k_y)^T$, where $R_1(k_x) = \exp(\frac{k_x+\pi}{2}L_z)$ and $R_2(k_y) = \exp[(k_y + \pi)L_y]$. The corresponding eigenstate pattern in a 2D first BZ is shown in Fig. 1(d). We see that the eigenstate frame rotates $\pi$ angle around the blue axis (3rd band) along $k_x: -\pi \to \pi$ for any fixed $k_y$. This rotation corresponds to the quaternion charge $+k$. Meanwhile, it also rotates $2\pi$ along $k_y: -\pi \to \pi$ for any fixed $k_x$, corresponding to charge $-1$. The 2D model can also be understood as a 2D subsystem (a torus surface) in the 3D parameter space as shown in Fig. 1(e). The torus surface corresponds to a 2D BZ. The poloidal and toroidal loops correspond to the $k_x$ and $k_y$ directions, respectively. They encircle a $+k$ nodal line inside the torus (blue ring) and a $-1$ nodal line (grey line) in the central void. The pattern in Fig. 1(d) can be obtained when we open up the torus into a rectangle. The 2D band structure is shown in Fig. 1(f). We note that the correspondence between the Bloch $k_x$ ($k_y$) and the loops on the toroidal surface can be exchanged so that we can also assign the poloidal loops as $k_y$ directions and the toroidal loops as $k_x$ directions. Then the nodal line configuration in the 3D parameter space will also be changed. However, it does not affect any properties of the 2D Hamiltonian defined on the torus surface.

From a homotopy point of view, we want to characterize the topological space of gapped Hamiltonians defined on the torus ($T^2$) of the BZ. For *N*-band PT symmetric systems, the second homotopy group $\pi_2(M_N) = 0$ is trivial [20]. Therefore, the classification of maps on the torus is equivalent to the one on the one-skeleton of the torus (bouquet of two circles) [29]. We can thus classify the system using an ordered pair as $(q_x, q_y)$, with each component being a quaternion defined over a 1D subsystem and we will call ordered pair a "2D non-Abelian charge". In our example, $(q_x, q_y) = (+k, -1)$, with the first and second component corresponding respectively to the 1D invariants along the $k_x$ and $k_y$ directions. Besides, the $q_x(q_y)$ is independent of the $k_y(k_x)$ since there are no degeneracies on the 2D BZ.

## B. Constraints on the 2D Zak phase and polarization

In literature, the 2D Zak phase and polarization are widely used for characterizing the topological properties of 2D PT symmetrical models [6]. The 2D Zak phase and polarization of the *n*-th band are defined as $\mathbf{Z_n} = \int dk_x\, dk_y\, \mathbf{A_n}(k_x, k_y)$ and $\mathbf{P_n} = \frac{\mathbf{Z_n}}{2\pi}$, respectively, where $\mathbf{A_n} = \langle \varphi_n | i\partial_k | \varphi_n \rangle$ is the Berry connection. The $\mathbf{Z_n}$ and $\mathbf{P_n}$ are two component vectors, and the two components correspond to *x* and *y* directions respectively. Here, we will show that the 2D non-Abelian charge pair defined above is very useful because it imposes some constraints on the 2D Zak phase and polarization. As an example, we will show that it is not possible for a 3-band system to exhibit 2D polarizations $(1/2, 0)$, $(1/2, 1/2)$ and $(0, 1/2)$ for the 1st, 2nd, 3rd band respectively.

To see why such configurations are not physically allowed, we construct a 2D model Hamiltonian in the form $H(k_x, k_y) = R_2(k_y)R_1(k_x)diag(-1,0,1)R_1(k_x)^T R_2(k_y)^T$, where $R_1(k_x) = \exp(\frac{k_x+\pi}{2}L_z)$ and $R_2(k_y) = \exp(\frac{k_y+\pi}{2}L_x)$. The eigenstate patterns of this system in $k$-space is shown in Fig. 1(g). We note that the eigenstates at $k_y = -\pi$ are different from the ones at $k_y = \pi$ indicating that the Bloch periodic condition cannot be satisfied. The constructed Hamiltonian has the property $H(k_x, -\pi) \neq H(k_x, \pi)$, so that the BZ boundary cannot be joined smoothly to form a torus. The situation can also be understood from an 3D viewpoint. We can always choose to regard this 2D Hamiltonian as a 2D subsystem in the 3D parameter space as shown in Fig. 1(h). In that higher dimension viewpoint, the 2D BZ is a torus embedded in 3D momentum space. As homotopy theories tell us that a rotating eigenvector frame on a $\pi_1$ loop encircles a degeneracy inside the loop, the rotating eigenvector frames in the 2D plane (as shown in Fig. 1(g)) implies that there is a nodal ring (blue) of quaternions $k$ inside the torus and a nodal line (red) of quaternions $i$ threading the central void as shown in Fig. 1(h). The 3 eigenstates on the torus will then be characterized by the 2D non-Abelian pair $(k, i)$ and the corresponding 2D single band polarizations are $p_x = (1/2, 1/2, 0)$ and $p_y = (0, 1/2, 1/2)$, with $(1/2, 1/2, 0)$ meaning that the eigenstates in 1st and 2nd bands are rotating while $(0, 1/2, 1/2)$ means that the 2nd and 3rd band eigenstates are rotating. The 3D picture can explain why such a configuration is not physical. According to the non-Abelian band topology [20], the reader is taken as the based point so that the orientation of the $+k$ nodal ring must change direction when it crosses under the $+i$ nodal line (marked by the blue triangle in Fig. 1(h)). The nodal line direction becomes opposite when it returns to the starting point (marked by the yellow point in Fig. 1(h)). The nodal lines with opposite orientations cannot annihilate each other. This corroborates with the picture shown in Fig. 1(g) that the edges of 2D square cannot be joined to form a torus, as the left and right edge eigen-frames are oriented differently. To make the system physically realizable, one possible configuration is shown in Fig. 1(k) where two additional nodal lines is emitted from this meeting point. These additional nodal lines will intersect the torus at 2 points, which are degeneracy points in the 2D subsystem as shown by the purple dots in Fig. 1(j). These two singularities will induce vortices which serves to make the left and right boundary configurations the same in Fig. 1(j) so that the flat BZ can be rolled up properly to become a torus. In short, the information embedded in quaternion (non-Abelian) topology can inform us that some 2D polarizations (e.g. Fig. 1(h)) are not physically realizable.

The two components in the 2D non-Abelian charge pair $(q_x, q_y)$ are required to be commutative, i.e., $q_x \cdot q_y = q_y \cdot q_x$ in order for the system to be a physically realizable periodic system. This can be understood as follows. In Fig. 1(i), we show two paths connecting the left-down corner to the right-up corner of the 2D BZ. The accumulated eigenstate rotations along the two paths correspond to $q_x \cdot q_y$ (brown path) and $q_y \cdot q_x$ (pink path), respectively. Since all the bands are required to be gapped, there are no degeneracies in the BZ. If the two paths can continuously transform from one to the other, the two rotations should produce the same result, indicating that $q_x \cdot q_y = q_y \cdot q_x$. However, quaternions are generally not commutative. If it happens that $q_x \cdot q_y \neq q_y \cdot q_x$, singularities (degeneracies) must exist inside the BZ. This is analogous to the Aharonov-Bohm effect, in which a flux tube must exist inside a loop if geometric phases accumulated along different paths connecting two points can be different. In the previous counter-example (Figs. 1(g)-1(h)), since $k \cdot i \neq i \cdot k$, the charge pair of type $(k, i)$ cannot exist in a periodic system that has no degeneracies inside the BZ.

We can construct a 2D model Hamiltonian with $q_x \cdot q_y \neq q_y \cdot q_x$ and carries degeneracies, corresponding to the 2D eigenvector frame patterns shown in Fig. 1(j) and the 3D degeneracy line configurations are shown in Fig. 1(k). The details of the Hamiltonian can be found in the section 2 of Ref. [28]. The corresponding band structure is shown in Figs. 1(l). To show the eigenstate pattern clearly, we also plot the eigenstate trajectories on the eigenstate sphere for several paths in Fig. S1 [28]. The two additional nodal lines characterized by the quaternion charge $k$ (blue lines in Fig. 1(k)) must penetrate the torus surface, resulting in singularities in the 2D eigenstate pattern (purple points in Fig. 1(j)) and degeneracy points in the band structure (Fig. 1(l)). Due to the degeneracies, we cannot define the 2D non-Abelian charge pair $(q_x, q_y)$ because the $q_x$ and $q_y$ are not constants in the 2D BZ now. In general, if the quaternion invariants $q_x$ and $q_y$ do not commute, the pair $(q_x, q_y)$ is not well defined because the order is important. The charges will change abruptly when the chosen loops meet the degeneracies. Therefore, the 2D single band polarization with $p_x = (1/2, 1/2, 0)$ and $p_y = (0, 1/2, 1/2)$ is forbidden in the gapped 3-band systems. The commutation relationships of the quaternion charges impose constraints on the 2D Zak phase and polarization. On the other hand, the configuration $(q_x, q_y) = (+k, -1)$, corresponding to Figs. 1(d) and 1(e) is allowed because the quaternion $-1$ commutes with the quaternion $k$.

## C. 3-band 2D non-Abelian topological insulator and the edge states

We now consider a 2D 3-band model which realizes the quaternion charges $(q_x, q_y) = (+k, -1)$ as shown in Figs. 1(d)-1(f). A tight-binding model that exhibits such topological character is shown pictorially in Fig. 2(a). There are 3 sites per unit cell. To be concise, each kind of hopping only shows up once and the full set of hopping can be envisioned by repeating the bonds periodically. The values of the hopping constants are listed in the section 2 of Ref. [28]. The tight-binding model is PT symmetric, and its Hamiltonian is real. We calculated the energy spectrum of a finite sample with 10 by 10 unit cells as shown in Fig. 2(b). The grey color indicates bulk states, while the red and blue are edge states localized on vertical (y-axis) and horizontal (x-axis) boundaries. The vertical edge states appear in the lower gap and the horizontal edge states appear in both of the two gaps. Figures 2(e)-2(g) show the representative field patterns of the vertical (Fig. 2(e)) and horizontal edge states (Figs. 2(f)-2(g)).

Then we calculate the eigenvalues of a vertical ribbon which is periodic along the y-direction (Fig. 2(c)) and the horizontal ribbon which is periodic along the x-direction

(Fig. 2(d)). The widths of the two ribbons are both 10 unit cells. For each fixed $k_y$ ($k_x$), the vertical (horizontal) ribbon Hamiltonian is equivalent to a 1D Hamiltonian and exhibits the 1D edge states. For the vertical ribbon, there is one edge state (per edge) in the lower gap between the 1st and 2nd band at each fixed $k_y$ (Fig. 2(c)), corresponding to quaternion $q_x = k$ each value of $k_y$. For the horizontal ribbon, the edge states correspond to the quaternion $q_y = -1$ each value of $k_x$. It is known that the bulk quaternion -1 can give rise to complicated edge state distributions as -1 can represent $i^2$ or $j^2$ or $k^2$ or their transition states. At each fixed $k_x$, the number of edge states (per edge) can be 2 or 3 and can pump between the gaps (see the section 3 of Ref. [28]). In Fig. 2(d), we see the edge states undergo the transition from those corresponding to the quaternion $j^2$ ($at\ k_x = -\pi$) to the $i^2$ ($at\ k_x = 0$) and go back to $j^2$ at $k_x = \pi$. The band structure of the ribbons is consistent with edge states of the finite sample in Fig. 2(b).

**D. 4-band 2D non-Abelian topological insulator and its edge states**

Now we move on to consider a 2D 4-band square lattice model, which is shown in Fig. 3(a). There are four sites, labelled as '1', '2', '3', '4' in a unit cell (dashed square). The thin and thick blue (red) bonds denote the intracellular and intercellular hopping along the $x$ ($y$) direction respectively. We label the intracellular hopping as $w_x$, $w_y$, and the intercellular hoppings as $v_x$, $v_y$, respectively. We first set $w_x = 1$, $v_x = 4$, $w_y = 2$, $v_y = 8$. The on-site energies are all set to zero. The momentum space Hamiltonian has the form:

$$H(k_x, k_y) = \begin{pmatrix} 0 & 0 & h_{13} & h_{14} \\ 0 & 0 & h_{14}^* & h_{13}^* \\ h_{13}^* & h_{14} & 0 & 0 \\ h_{14}^* & h_{13} & 0 & 0 \end{pmatrix}, \quad (1)$$

where $h_{13} = w_y + v_y \exp(ik_y)$, $h_{14} = w_x + v_x \exp(-ik_x)$. Here we have set the lattice constant as unity for convenience. The corresponding 2D band structures are shown in Fig. 3(b). We also calculated the band structures for finite width ribbons aligned along $y$ and $x$ directions as shown in Figs. 3(c) and 3(d), respectively. The ribbon in Fig. 3(c) (Fig. 3(d)) has 10 unit cells along the $x$-direction ($y$-direction) and is infinitely long along the $y$-direction ($x$-direction) so that the $k_y$ ($k_x$) is well defined. For each fixed $k_y$ ($k_x$), the vertical (horizontal) ribbon Hamiltonian is equivalent to a 1D subsystem and has a set of edge states. The edge states of the vertical ribbon (red curves in Fig. 3(c)) are found in the gaps of 1-2 and 3-4 bands while the ones of the horizontal ribbon (blue curves in Fig. 3(d)) overlap with the 2nd and 3rd bulk band continuum. We further calculated the energy spectrum of a finite lattice with $10 \times 10$ unit cells as shown in Fig. 3(e). The grey states are the bulk states. The red and blue states are edge states localized respectively at the vertical and horizontal edges. The yellow states are the corner states. The representative eigenstate patterns of the edge states and corner states are shown in Fig. 3(f)-(h), respectively. The inset of Fig. 3(e) shows that there are 4 corner states in total.

Now we explain the edge and corner state distributions from a multiband perspective. The four bands of the 2D band structure in Fig. 3(b) are gapped so that the 2D non-Abelian charge can be well defined. Before we calculate the 2D non-Abelian charge pair, we need to transform the Hamiltonian into a real Hamiltonian by a similarity transformation: $\widetilde{H}(\boldsymbol{k}) = V \cdot H(\boldsymbol{k}) \cdot V^\dagger$, where

$$V = \frac{1}{2}\begin{pmatrix} 1 & 1 & -i & i \\ 1 & 1 & i & -i \\ -i & i & 1 & 1 \\ -i & i & -1 & -1 \end{pmatrix}. \tag{2}$$

Then we can obtain a real Hamiltonian:

$$H(k_x, k_y) = \begin{pmatrix} s_1 & 0 & t_{13} & t_{14} \\ 0 & s_2 & -t_{14} & -t_{13} \\ t_{13} & -t_{14} & s_3 & 0 \\ t_{14} & -t_{13} & 0 & s_4 \end{pmatrix}. \tag{3}$$

Here, $t_{13} = w_y + v_y\cos(k_y)$, $t_{14} = -w_x - v_x\cos(k_x)$, $s_1 = -v_x\sin(k_x) - v_y\sin(k_y)$, $s_2 = v_x\sin(k_x) + v_y\sin(k_y)$, $s_3 = -v_x\sin(k_x) + v_y\sin(k_y)$, $s_4 =$

$v_x \sin(k_x) - v_y \sin(k_y)$. All the matrix elements are real numbers in this basis, and hence the eigenvectors are real eigenvectors $(u_1, u_2, u_3, u_4)^T$. The four real number components compose a 4D parameter space and the eigenvectors can be described by the trajectories in this 4D space as $k_x$ or $k_y$ changes in momentum space.

According to Ref. [24], the non-Abelian group describing the eigenvector frame rotation in a 4-band system is the generalized quaternion group $Q_{16}$, and it corresponds to a 4D rotation in the 4D parameter space. The group $Q_{16}$ has 16 elements in 10 conjugacy classes: $\{+1\}$, $\{-1\}$, $\{\pm q_{12}\}$, $\{\pm q_{13}\}$, $\{\pm q_{14}\}$, $\{\pm q_{23}\}$, $\{\pm q_{24}\}$, $\{\pm q_{34}\}$, $\{+q_{1234}\}$, $\{-q_{1234}\}$. The $\{+1\}$ is topologically "trivial" and the $\{-1\}$ class corresponds to a $2\pi$ simple frame rotation on any rotation plane. The $\{\pm q_{mn}\}$ correspond to a $\pm\pi$ simple rotation on the rotation plane spanned by the eigenvectors of the *m*-th and *n*-th bands. The $\{+q_{1234}\}$ and $\{-q_{1234}\}$ correspond to the left and right isoclinic rotations of angle $\pi$, respectively. The isoclinic rotation is a special double rotation that the rotation angles are the same on the two orthogonal rotation planes. It can be further divided into the left and right isoclinic rotations by the sign difference on the two rotation planes. For example, the charge $+q_{1234}$ can be combined by the $+q_{12}$ and $+q_{34}$, while the charge $-q_{1234}$ can be combined by $-q_{12}$ and $+q_{34}$. We note that the charges $+q_{1234}$ and $-q_{1234}$ belong to two different conjugacy classes as they cannot continuously transform to each other with gap closing even without a based point.

We calculated the 2D non-Abelian charge pair for the model in Fig. 3(a) and found $(q_x, q_y) = (q_{1234}, -q_{1234})$. The detailed calculation of the 4-band non-Abelian charge can be found in the section 1 of Ref. [28]. In the first panel of Fig. 3(i), the green arrow indicates the path with $k_y = 0$ and $k_x$ running from $-\pi$ to $\pi$ in the BZ. As the $q_x$ is found to be $q_{1234}$, the eigenstate rotations along this path should correspond to the left isoclinic rotation of $\pi$. The eigenstate trajectories along this path are shown in the latter four panels in Fig. 3(i). To see the 4D eigenstate rotations explicitly, we adopt the orthographical projection method. In this projection, the 4D trajectories are projected into four 3D subspaces from four orthogonal views. This is similar to the method of three-view drawing, which is the orthographic projection from 3D space to 2D plane. Take the first projection of Fig. 3(i) as an example, the trajectories in the $u_1u_2u_3$ subspace are plotted so that it is an orthographic projection from the view of $u_4$ direction. The red, cyan, magenta, blue trajectories correspond to the eigenstates of the 1st, 2nd, 3rd, 4th bands respectively. The increasing thickness of the line indicates the direction of $k_x$ running from $-\pi$ to $\pi$. From the projections, we see that the eigenstates of the 1st and 2nd bands rotate $-\pi$ in the same plane, and the eigenstates of the 3rd and 4th bands rotate $-\pi$ in another orthogonal plane. Therefore, the $q_x = q_{1234}$ is the combination of $(-q_{12})(-q_{34}) = q_{12}q_{34}$ at $k_y = 0$. The $q_{1234}$ $(-q_{1234})$ charge can be factorized in three ways: $q_{12}q_{34}$ $(-q_{12}q_{34})$, $-q_{13}q_{24}$ $(q_{13}q_{24})$ and $q_{14}q_{23}$ $(-q_{14}q_{23})$ [24]. At the hard boundary of a 1D system, these three configurations have different edge state distributions in the band gaps (see the section 4 of Ref. [28]). For the factorizations $\pm q_{kl}q_{mn}$, one edge state (per edge) is found between the *k*-th and *l*-th bands, and the other edge state (per edge) is between the *m*-th and *n*-th bands. In our case, we will show in the next section that the $q_x$ is always the factorization of $q_{12}q_{34}$ for any $k_y$. The charge $q_x = q_{12}q_{34}$ indicates that the edge states along the vertical edges should exist in the gaps of 1-2 bands and 3-4 bands respectively. The band structure in Fig. 3(c) is consistent with this indication.

In the same way, we study the non-Abelian charge along the *y*-direction. In Fig. 3(j), we plot the eigenstate rotations along the path with $k_x = 0$ with $k_y$ running from $-\pi$ to $\pi$ (green arrow). It can be seen from the eigenvector rotations that the $q_y = -q_{1234}$ is the combination of $q_{13}$ and $q_{24}$ at $k_x = 0$. For other values of $k_x$, we also find that the $q_y$ is always the factorization of $q_{13}q_{24}$ at any $k_x$ (explained in the next section). The 1-3 band pair and 2-4 pair of bands are always the two rotation planes for any $k_x$ as $k_y$ traverses the BZ. The charge $q_y = q_{13}q_{24}$ indicates that the edge states along the horizontal edges should exist in the band gaps of 1-3 bands and 2-4 bands respectively. The band structure of the horizontal ribbon in Fig. 3(d) verifies this prediction. The blue curves in Fig. 3(d) are the horizontal edge states and they are sandwiched between the 1-3 bands and 2-4 bands. We note the edge mode between the 1-3 bands overlaps with the 2nd bulk band continuum and the edge mode between the

2-4 bands overlaps with the 3rd bulk band. As the two pairs of bands (1-3 pair and 2-4 pair) are entirely decoupled (explained in the next section), the overlapped edge states and bulk states do not interact, and the edge modes can exist as bound states in the bulk continuum.

### E. Off-diagonal Berry connection and off-diagonal Berry phase

In the model mentioned above, the non-Abelian charges $q_x$ and $q_y$ are represented respectively by the factorizations $q_{12}q_{34}$ and $q_{13}q_{24}$. For more general cases, however, the non-Abelian charge $q_{1234}$ (or $-q_{1234}$) are the hybrids of the three factorizations: $q_{12}q_{34}$, $-q_{13}q_{24}$, $q_{14}q_{23}$ [24]. The three factorizations belong to the same non-Abelian charge, and they can evolve to each other without gap closing. We cannot determine the factorization information only from the non-Abelian charges.

To solve this problem, we propose to add the Euler connection [21] (also called off-diagonal Berry connection [30]) as additional information to characterize the eigenstate rotations. The Euler connection $\boldsymbol{A}^{mn}(\boldsymbol{k})$ is defined as $\boldsymbol{A}^{mn}(\boldsymbol{k}) = \langle \varphi^m(\boldsymbol{k}) | \nabla_{\boldsymbol{k}} | \varphi^n(\boldsymbol{k}) \rangle$ [21]. Here all the $|\varphi^j(\boldsymbol{k})\rangle$ are the real eigenstates. The geometrical meaning of the $\boldsymbol{A}^{mn}(\boldsymbol{k})$ is the instantaneous projected rotation angle of the $|\varphi^m(\boldsymbol{k})\rangle$ (or $|\varphi^n(\boldsymbol{k})\rangle$) on the *m-n* plane as shown in Fig. 3(k). Then we integrate the Euler connection along a closed loop on the BZ:

$$\gamma^{mn} = \oint \boldsymbol{A}^{mn}(\boldsymbol{k}) \cdot d\boldsymbol{k}. \qquad (4)$$

The sign of $\gamma^{mn}$ indicates the rotation direction. It is referred to as the off-diagonal Berry phase (OBP) in literature [30]. We note that the OBP is not necessarily a quantized number. It is quantized only if the rotation plane is strictly fixed in the *m-n* plane (in ideal models).

We can further write down the OBP matrix $\boldsymbol{\gamma}$ whose *m-n* element is the OBP $\gamma^{mn}$. The OBP matrix is a real skew-Hermitian matrix that $\gamma^{mn} = -\gamma^{nm}$ and all the elements are real numbers. As we want to know the relations between different bands, we only need to focus on the off-diagonal terms. For an ideal model in which the *m-n* plane is always the rotation plane, the total rotation angle of the $|\varphi^m(\boldsymbol{k})\rangle$ (or $|\varphi^n(\boldsymbol{k})\rangle$) on the *m-n* plane is a quantized number $\gamma^{mn,m>n}$. Here we set the positive sign of the elements under the diagonal elements ($m > n$) as the positive rotation direction. Then there should be $|\gamma^{mn}|/\pi$ edge states (per edge) between the *m*-th and *n*-th bands in the ideal model. For general cases, the *m-n* plane is not necessarily the rotation plane so the $|\gamma^{mn}|/\pi$ is not necessarily an integer.

Now we show the calculated results for the OBP matrix elements for the paths in the 2D BZ in Fig. 3(i) and 3(j) respectively. The matrix is shown as follows.

$$\boldsymbol{\gamma}_x(k_y = 0) = \begin{pmatrix} \pi & \pi & 0 & 0 \\ -\pi & \pi & 0 & 0 \\ 0 & 0 & \pi & \pi \\ 0 & 0 & -\pi & \pi \end{pmatrix}, \tag{5}$$

and

$$\boldsymbol{\gamma}_y(k_x = 0) = \begin{pmatrix} \pi & 0 & -\pi & 0 \\ 0 & \pi & 0 & -\pi \\ \pi & 0 & \pi & 0 \\ 0 & \pi & 0 & \pi \end{pmatrix}. \tag{6}$$

Here, the $\boldsymbol{\gamma}_x$ and $\boldsymbol{\gamma}_y$ represent the OBP matrices integrated along the *x* and *y* direction, respectively. The diagonal Berry phases are calculated using the phase differences between the first and last eigenstates $\gamma^{mm} = \text{Im}[\ln\langle\varphi^m(\boldsymbol{k}_0)|\varphi^m(\boldsymbol{k}_N)\rangle]$. For the $\boldsymbol{\gamma}_x(k_y = 0)$, the $\gamma_x^{21}$ and $\gamma_x^{43}$ are both $-\pi$ indicating that the eigenstate rotation is the $-\pi$ double rotation on the plane spanned by 1st and 2nd eigenvectors and plane spanned by the 3rd and 4th eigenvectors respectively. The corresponding non-Abelian charge factorization is $q_{1234} = (-q_{12})(-q_{34})$, which is consistent with the 4D rotation in Fig. 3(i). For the $\boldsymbol{\gamma}_y(k_x = 0)$, the $\gamma_x^{31}$ and $\gamma_x^{42}$ are both $\pi$, indicating that the $-q_{1234} = q_{13}q_{24}$. This is also consistent with the 4D rotation shown in Fig. 3(j).

To show that the $q_x$ is always the factorization of $(-q_{12})(-q_{34})$ for any $k_y$ in our model, we calculate the $\gamma_x^{mn}(k_y)$ and results are shown in Fig. 3(l). The $\gamma_x^{21}$ and $\gamma_x^{43}$ are always $-\pi$ while other elements are always zeros for all the $k_y$, indicating that $q_x(k_y: -\pi \to \pi) = (-q_{12})(-q_{34})$. For the other direction, the $\gamma_y^{mn}(k_x)$ curves are shown in Fig. 3(m). The $\gamma_y^{31}$ and $\gamma_y^{42}$ are always $\pi$ while other elements are always zeros for all the $k_x$ indicating that $q_y(k_x: -\pi \to \pi) = q_{13}q_{24}$. The 1-3 and 2-4 band pairs are entirely decoupled for any $k_x$ so that they can be viewed as two separated 2-band Hamiltonians under the certain transformation. It explains why the vertical edge states can overlap with the continuum of 2$^{nd}$ and 3$^{rd}$ bulk bands in the form of bound states in continuum (BICs). From the analysis above, we see that the $\gamma_x^{mn}(k_y)$ and $\gamma_y^{mn}(k_x)$ curves can provide useful information about the couplings between the bands. They can help us to differentiate the three factorizations of $\pm q_{1234}$ and hence provide sufficient information to predict the edge state distributions, such as the ribbon band structures in Fig. 3(c) and 3(d).

### F. Corner states originating from edge states

Using the notion of 2D non-Abelian charge pair and the OBP matrix information, we predict the edge state distributions in the energy spectrum. Now we explain the topological origin of the corner states from this multiband perspective. Considering the vertical ribbon in Fig. 3(c), we can represent its vertical edge state as a $k_y$-dependent wave function as $|\varphi_{L,R}^m(k_y)\rangle$, where the subscripts $L$ and $R$ denote the left and right edges respectively, and the superscripts $m = 1$ or $2$ denote the band indices of the edge states. Since there are only two edge state bands, we can use the Berry phase to characterize its topological properties individually. Taking the 1$^{st}$ band of the left edge state as an example, we can calculate its Berry phase by the continued multiplication along $k_y$ direction:

$$\phi = \text{Im}[\ln\langle\varphi_L^1(k_0)|\varphi_L^1(k_1)\rangle\langle\varphi_L^1(k_1)|\varphi_L^1(k_2)\rangle\ldots\langle\varphi_L^1(k_{N-1})|\varphi_L^1(k_0)\rangle]. \quad (7)$$

We get the Berry phases $\pi$ for the vertical edge states as labeled in Fig. 3(c). If we cut the vertical ribbon into a finite length, the quantized Berry phase $\pi$ implies the existence of the 0D edge states in the gap between the two edge state bands, which are the corner states. The corner states in this system can hence be viewed as the 0D edge states of the 1D edge states if they fall in the gap of the bulk continuum. As each edge state (left and right) supports one boundary mode at one end, there are a total of four corner states. We can also change the cutting sequence that we first cut the infinitely large lattice into a horizontal ribbon, then cut the horizontal ribbon into a finite length. Then we also get the Berry phases $\pi$ for the horizontal edge states as labeled in Fig. 3(d). The two cutting sequences consistently predict the same corner state distribution.

## G. Predicting edge states and corner states using 2D Zak phases

In the above discussion, we proposed using the 2D non-Abelian band topology to understand the topological origins of the edge and corner states. However, one may wonder why we use a multi-band non-Abelian theory to study simple models such as the 2D SSH lattice. To show the usefulness of the 2D multi-band theory, here we analyze the previous models with the widely used standard 2D Zak phase method and we will find that it is not always correct in predicting the edge and corner states.

In literature, the 2D Zak phase is related to the edge and corner states [8] in a gap for the rectangle samples. The 2D Zak phase of a target gap is calculated by summing up (vector addition) the 2D Zak phases of all the bands below this gap. The $(0,0)$ is the trivial case. The $(0,\pi)$ and $(\pi,0)$ cases imply the edge states in only one direction. A bulk Zak phase of $(\pi,\pi)$ implies the existence of edge states in both directions and also the corner states. Now we go back to the 3-band model in Fig. 2. The 2D Zak phases of the individual bands are $(\pi,0)$, $(\pi,0)$ and $(0,0)$ for the 1$^{st}$, 2$^{nd}$ and 3$^{rd}$ bands respectively. The corresponding 2D Zak phases of the 1$^{st}$ and 2$^{nd}$ gaps are $(\pi,0)$ and $(0,0)$ respectively. According to the Zak phases, there is only one group of vertical edge states in the first gap. However, there are horizontal edge states in both gaps which are beyond the prediction of the 2D Zak phase.

Then we move to the 4-band model in Fig. 3. The 2D Zak phases of the individual bands are $(\pi,\pi)$, $(\pi,\pi)$, $(\pi,\pi)$ and $(\pi,\pi)$ for the 1$^{st}$, 2$^{nd}$, 3$^{rd}$ and 4$^{th}$ bands respectively. The 2D Zak phase of the 1$^{st}$, 2$^{nd}$ and 3$^{rd}$ gaps are $(\pi,\pi)$, $(0,0)$ and $(\pi,\pi)$ respectively. This should imply edge states in both directions and the corner states in the 1$^{st}$ and 3$^{rd}$ gaps, and there should be no edge states and corner states in the 2$^{nd}$ gap. However, the fact is that there are only vertical edge states in the 1$^{st}$ and 3$^{rd}$ gaps and the corner states are in the 2$^{nd}$ gap which is thought to be topologically trivial. Using the two examples, we demonstrate that the 2D Zak phases may not be always reliable in predicting the edge and corner states. The non-Abelian band topology can explain the topological origins of the edge and corner states in those examples.

### III. A MORE GENERALIZED MODEL WITH DEGENERACY POINTS

In the above discussion, the 2D non-Abelian topological insulator requires the gapped bands. In this section, we will extend the consideration to PT symmetrical systems that carry degeneracy points. A prototypical tight-binding model is shown in Fig. 4(a). Different from the one in Fig. 3(a), the onsite energies of the model in Fig. 4(a) can be tuned and second neighbor diagonal (anti-diagonal) hopping terms are added. The onsite energies of sublattice sites 1, 2 are set as $S_A = 2$ and those of sublattice sites 3, 4 are set as $S_B = -2$. The intra-cell and inter-cell diagonal hopping terms are set as $\kappa_1 = 1$ and $\kappa_2 = 3$. The intra-cell and inter-cell anti-diagonal hoppings are set as $\lambda_1 = 2$ and $\lambda_2 = 6$. The $w_x$, $v_x$, $w_y$, $v_y$ are set as $w_x = 1.8$, $v_x = 4.5$, $w_y = 0.8$, $v_y = 1.7$.

The corresponding 2D band structure is shown in Fig. 4(b). Owing to the additional diagonal and anti-diagonal hopping terms, there are two linear degeneracy points between the 2$^{nd}$ and 3$^{rd}$ bands located at $(k_x, k_y) = (-2.513, -1.068)$ and (2.513,1.068), respectively. Due to the degeneracies, we cannot define an invariant 2D non-Abelian charge $(q_x, q_y)$. However, for those paths that do not cross the degeneracies in the BZ, we can still calculate the $k_y$ and $k_x$ dependent non-Abelian charges along the x and y directions i.e., $q_x(k_y)$ and $q_y(k_x)$.

In the top panel of Fig. 4(d), we show the calculated $q_y(k_x)$ for different domains in the BZ. The two degeneracy points divide the BZ into three parts with $k_x$ going from left to right. The corresponding non-Abelian charges $q_y(k_x)$ are found to be $-q_{14}$, $-q_{1234}$, $-q_{14}$. In the same way, the charges $q_x(k_y)$ are found to be $-q_{14}$, $q_{1234}$, $-q_{14}$ for $k_y$ from top to bottom domains as shown in the left panel in Fig. 4(e). If we extend to a higher dimension view as shown in Fig. 4(c), the topological charges have a corresponding nodal line configuration, in which the 2D system BZ is subsystem represented by a torus. Two $q_{23}$ nodal lines punch through the torus surface at two degeneracy points between the 2nd and 3rd bands on the 2D BZ.

Furthermore, the OBPs are also well defined for the paths not crossing the degeneracies. In the middle panel of Fig. 4(d), we calculate the $\gamma_y^{mn}(k_x)$ curves for different domains. It can be seen that the $\gamma_y^{mn}(k_x)$ curves have sudden jumps when they cross the degeneracy points. The edge state distributions of a horizontal ribbon (as shown in the bottom panel in Fig. 4(d)) can be inferred from the information of the non-Abelian charges $q_y(k_x)$ and $\gamma_y^{mn}(k_x)$ curves. The ribbon has 10 unit cells along the y-direction and infinitely long along the x-direction.

For the two $q_y(k_x) = -q_{14}$ domains, the $\gamma_y^{21}(k_x)$, $\gamma_y^{43}(k_x)$, $\gamma_y^{32}(k_x)$ and $\gamma_y^{41}(k_x)$ are nonzero. According to Ref. [24], there are usually two possible configurations for $\pm q_{14}$ (see the section 4 of Ref. [28].). One allowed configuration is very special that the eigenstate rotation plane is always the 1-4 plane and the 1-4 band pair can be decoupled from other bands. In this case, the $\gamma_y^{21}(k_x)$, $\gamma_y^{32}(k_x)$, $\gamma_y^{43}(k_x)$ should all be zero and there is only one edge state (per edge) sandwiched between the 1st and the 4th band. The other configuration is the general case that all the four bands are coupled together. In this case, the $\gamma_y^{21}(k_x)$, $\gamma_y^{32}(k_x)$, $\gamma_y^{43}(k_x)$ are nonzero and each gap supports one edge state (per edge) so that there are three edge states (per edge) in total. We also show the transition between the two cases in Fig. S3(a) [28]. As long as the coupling between the 1-4 band pair and other bands are nonzero, two extra edge states (per edge) will emerge from the bulk bands. In our case, the non-zero $\gamma_y^{21}(k_x)$, $\gamma_y^{32}(k_x)$, $\gamma_y^{43}(k_x)$ curves indicate that it is the general case. As shown in the bottom panel of Fig. 4(d), the edge state distributions (blue curves) in the two $-q_{14}$ domains are consistent that each gap supports one edge state per edge. Due to the finite size effect, the edge states in the 2-3 gap do not fill in the whole $-q_{14}$ domains. If we increase the unit cell number along the y-direction, the ends of the edge state curve will converge to the degeneracy points and this curve will finally connect the two degeneracy points.

For the $q_y(k_x) = -q_{1234}$ domain, we see $\gamma_y^{21}(k_x)$ and $\gamma_y^{43}(k_x)$ curves have bigger absolute values indicating that these $-q_{1234}$ charges can be approximately factorized as the $-q_{12}q_{34}$. This is consistent with the bottom panel of Fig. 4(d), in which the edge states of the $-q_{1234}$ domain are in the 1-2 and 3-4 gaps. Through a similar analysis, we can explain the edge state distributions of a vertical ribbon from the $q_x(k_y)$ and $\gamma_x^{mn}(k_y)$ curves as shown in Fig. 4(e).

Then we calculate the energy spectrum of a 10 by 10 finite sample as shown in Fig. 4(f). The grey, red and blue states correspond to bulk, vertical edge and horizontal edge states, respectively. The energy distributions of edge states of this square sample are consistent with those of the two ribbons in Figs. 4(d) and 4(e). We note that some edge states in Fig. 4(f) overlap with the bulk bands. By calculating the complex energy spectrum for the sample with background loss [16], we find that they have couplings with bulk states. So straightly speaking, they are surface resonances. Figures 4(g)-4(h) show the examples of the vertical edge and horizontal edge states respectively.

## IV.   CONCLUSION

In this study, we consider 2D topological insulators from a multi-band non-Abelian band perspective. The non-Abelian charge concept imposes constraints on the 2D polarization. We applied the non-Abelian topological theory to 2D 3-band and 4-band model systems and explained the topological origin of the edge states and corner states from a multi-band perspective. The existence of edge states can be explained by non-trivial 2D quaternion charges, and the corner states are the 0D boundary modes of the 1D edge states. Furthermore, we noted that the off-diagonal Berry phase can provide complementary information. Knowing the 2D quaternion charge pair and the OBP, we can predict the energy distribution of the edge states in situations where the 2D Zak phase method fails. Compared with the 2D Zak phase, the non-Abelian charge method can provide more information by considering multiple bands as a whole.

**Acknowledgements**

This work is supported by the Hong Kong RGC (16307821, AoE/P-502/20), the KAUST CRG grant (KAUST20SC01) and the Croucher Foundation (CAS20SC01).

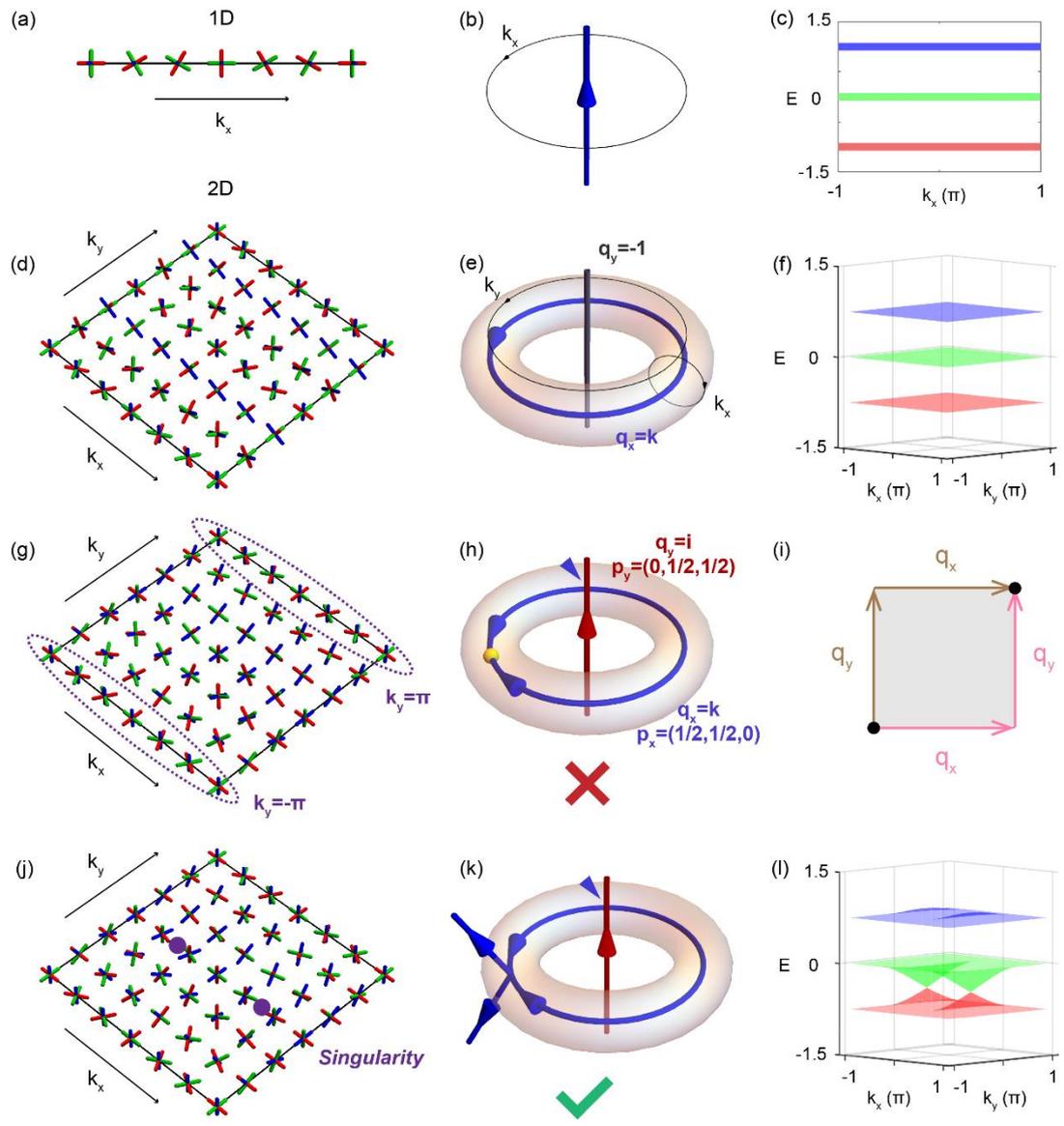

FIG. 1. Non-Abelian topological insulator. (a) The eigenstate rotations in a 1D non-Abelian topological insulator. (b) A 1D non-Abelian topological insulator can be viewed as a subsystem which is a closed loop encircling nodal lines (degeneracies) in a higher dimension parameter space. (c) The 1D band structure of a flat band model. (d) The eigenstate pattern in a 2D non-Abelian topological insulator. (e) A 2D non-Abelian topological insulator can be viewed as a torus surface encircling nodal lines in 3D. (f) The 2D band structure of a flat band model. (g) The eigenstate pattern which does not satisfy Bloch condition. (h) Forbidden nodal line configurations impose constraints on the 2D polarizations. (i) Two paths in a 2D BZ. (j) The eigenstate pattern with singularities. (k) An admissible nodal line configuration with two nodal lines penetrating the torus surface. (l) The band structure with degeneracies.

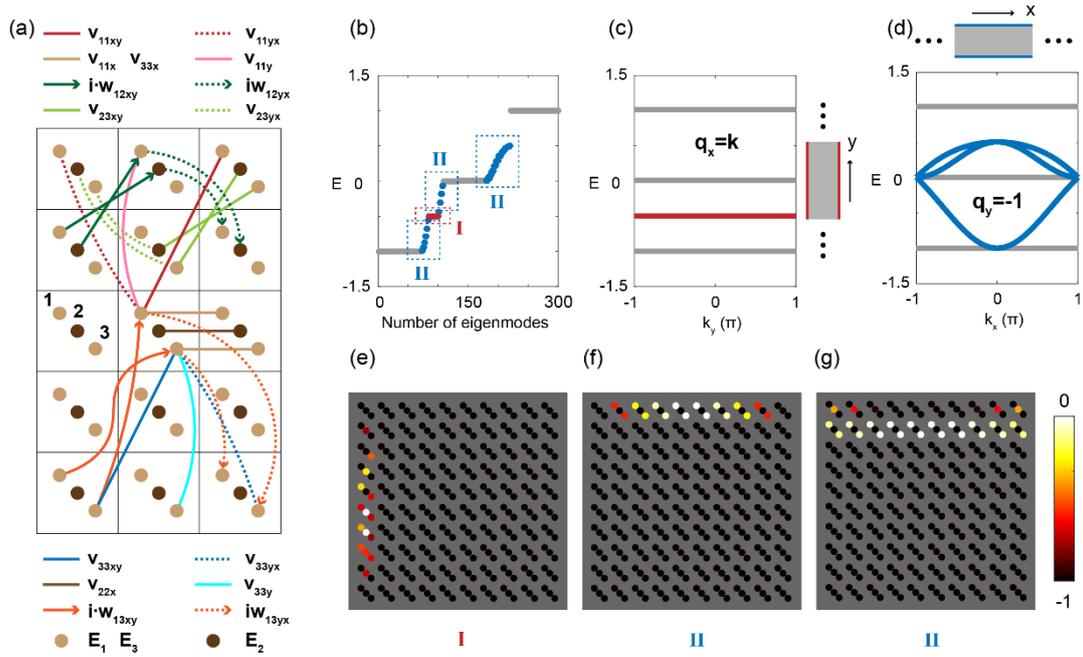

FIG. 2. 3-band 2D non-Abelian topological insulator. (a) Tight-binding model. (b) The energy spectrum of a finite sample with 10 by 10 unit cells. The grey, red and blue states are bulk, vertical and horizontal edge states respectively. (c) The band structure of a vertical ribbon. (d) The band structure of a horizontal ribbon. (e)-(g) The representative field patterns of vertical (e) and horizontal edge states (f, g).

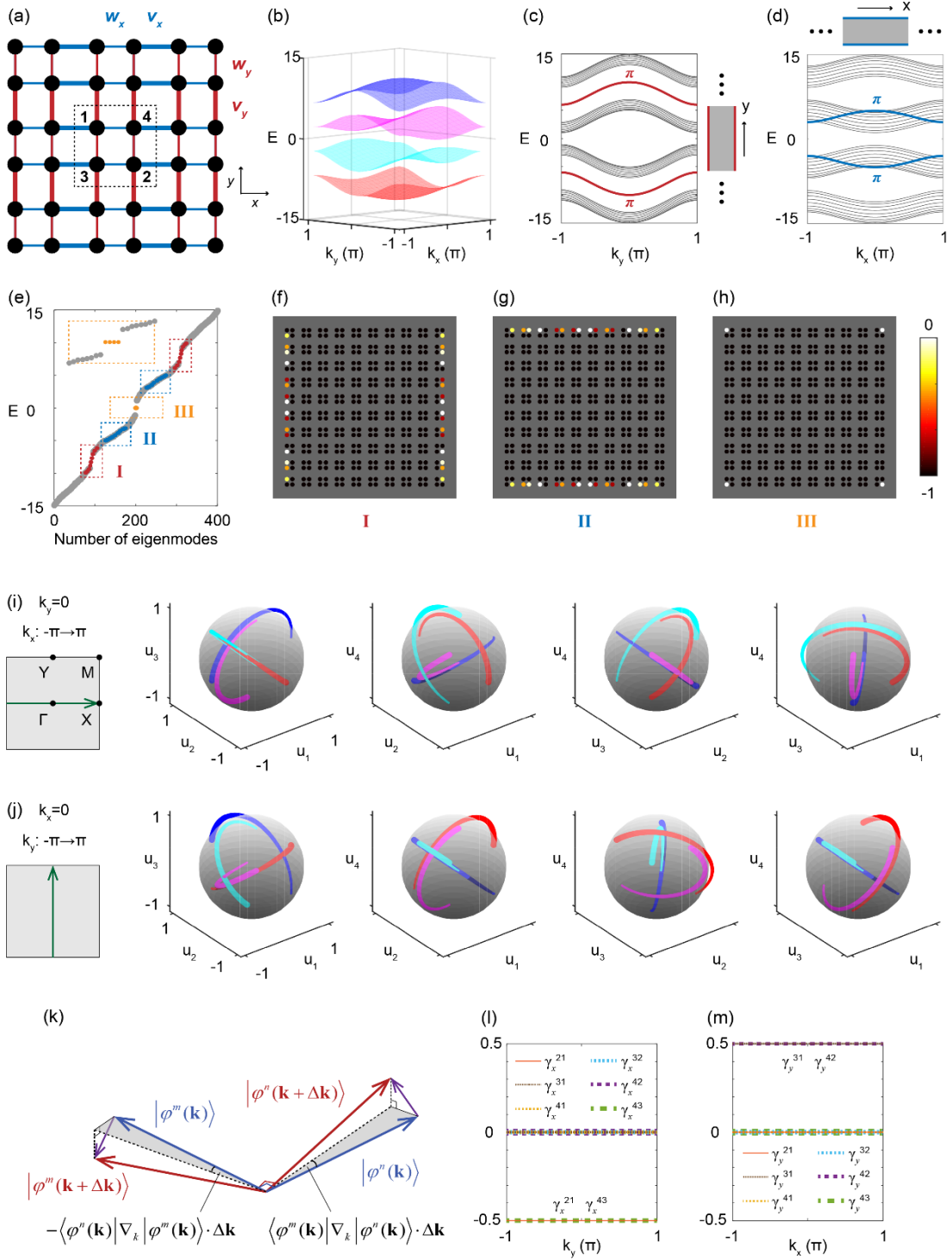

FIG. 3. 4-band 2D non-Abelian topological insulator. (a) The tight-binding model. (b) 2D band structure. (c)-(d) The band structures of a vertical ribbon and a horizontal ribbon respectively. (e) The energy spectrum of a $10 \times 10$ lattice. (f)-(h) Examples of the strength field patterns of vertical edge, horizontal edge and corner states respectively. (i) The orthographical projections of the eigenstate rotations along the path $k_y = 0$, $k_x: -\pi \to \pi$. (j) The orthographical projections of the eigenstate rotations along the path $k_x = 0$, $k_y: -\pi \to \pi$. (k) The geometrical meaning of the Euler connection. The shaded area is the projected area scanned by the $|\varphi^m(\boldsymbol{k})\rangle$ (or $|\varphi^n(\boldsymbol{k})\rangle$) on the $m$-$n$ plane in the $\Delta \boldsymbol{k}$. (l) The OBP curves $\gamma_x^{mn}(k_y)$. (m) The OBP curves $\gamma_y^{mn}(k_x)$.

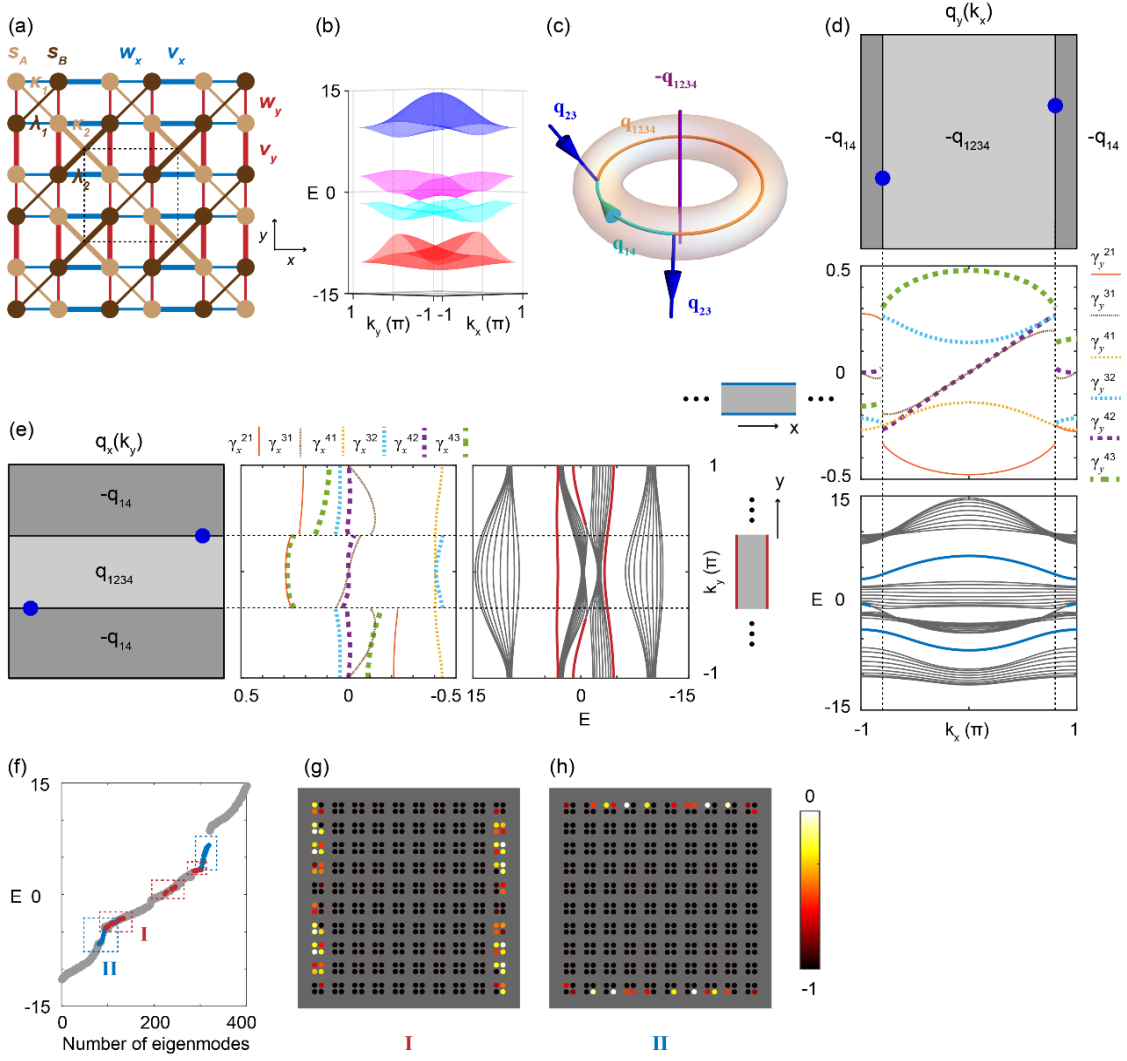

FIG. 4. A model with degeneracy points. (a) The tight-binding model with the diagonal and anti-diagonal hopping. (b) The 2D band structure. There are two degeneracy points between the 2nd and 3rd bands. (c) The corresponding nodal line configuration viewed from higher dimensions. (d) Top panel: the non-Abelian charges $q_y(k_x)$ in different subdomains of the 2D BZ. Middle panel: The Euler phase curves $\gamma_y^{mn}(k_x)$. Bottom panel: The band structure of a horizontal ribbon (see inset). (e) Left panel: the non-Abelian charges $q_x(k_y)$. Middle panel: The Euler phase curves $\gamma_y^{mn}(k_x)$. Right panel: The band structure of a vertical ribbon (see inset). (f) The energy spectrum of the $10 \times 10$ lattice. (g)-(h) Examples of the vertical edge, horizontal edge states respectively.